\documentclass{article}
\usepackage[hang]{footmisc}
\usepackage{footmisc,multicol, graphicx,makeidx,courier,helvet}
\usepackage[latin1]{inputenc}
\usepackage{amssymb,amsmath,amsthm} 
\usepackage{amsfonts}
\usepackage{graphicx}

\begin{document}

\title{Towards a new brain science: lessons from the economic collapse}

\author{Jaime Gomez-Ramirez\footnotemark[1],\footnotemark[2] and Manuel G.
Bedia \footnotemark[3]} \footnotetext[1]{Universidad Politécnica de
Madrid\newline \hspace{25pt}José Gutiérrez Abascal, 2 Madrid 28006  
\newline \emph{E-mail address}:jd.gomez@upm.es}
\footnotetext[2]{ Biomedical Engineering Laboratory, Okayama
University\newline \hspace{25pt}700-8530 1-1-1 Tsushima-naka , Kita-ku , Okayama-shi  
}
\footnotetext[3]{Universidad de Zaragoza\newline
\hspace{25pt}María de Luna 1, 50018 Zaragoza 
\newline \emph{E-mail address}:mgbedia@unizar.es}

\date{}
\maketitle
\abstract{
Economies are complex man-made systems where organisms and markets interact
according to motivations and principles not entirely understood yet. The
increasing dissatisfaction with the postulates of traditional economics i.e. perfectly rational agents,
interacting through efficient markets in the search of equilibrium, has
created new incentives for different approaches in economics. The science of
complexity may provide the platform to cross disciplinary boundaries in
seemingly disparate fields such as brain science and economics. In this paper
we take an integrative stance, fostering new insights into the economic
character of neural activity. The objective here is to precisely delineate
common topics in both neural and economic science, within a systemic outlook
grounded in empirical basis that jolts the unification across the science of
complex systems. It is argued that this mainly relies on the study of the
inverse problem in complex system with a truly Bayesian approach.}

\section{Introduction}
Since the financial crash in 2008, economic science and the economic profession
are under siege. Critics point fingers at ivory tower economists, devoted to
the construction  of unfalsifiable models based on unrealistic
assumptions in purely theoretical basis.
Economies are complex man-made systems where organisms and markets interact
according to motivations and principles not entirely understood yet.
Neo-classical economics is agnostic about the neural mechanisms that underlie
the valuation of choices and decision making. The increasing dissatisfaction with 
the postulates of traditional economics i.e. perfectly rational agents,
interacting through efficient markets in the search of equilibrium, has
created new incentives for different approaches in economics. Behavioral
economics \cite{Akerlof:2002},\cite{Smith:2009} builds on cognitive
and emotional models of agents, neuroeconomics addresses the neurobiological
basis of valuation of choices \cite{Montague:2002},\cite{Glimcher:2008} or
Evolutionary economics \cite{Anderson:1989},
\cite{Arthur:1997},\cite{Arthur:1999},\cite{Arthur:1999b},\cite{Blume:2005}
which strives for a new understanding of the economy as a complex evolutionary
system, composed of agents that adapt to endogenous patterns out of
equilibrium regions. The science of complexity may provide the platform to
cross disciplinary boundaries in seemingly disparate fields such as brain
science and economics. 
Social science, and in particular economics, is undergoing a decisive
historical moment. New mathematical
models able to palliate the dissatisfaction with core tenets in classical economics, like
rational agents, symmetric information and equilibrium need to be devised.

We argue that the most important problems that natural and social science are
facing today are inverse problems, and that a new approach that goes beyond
optimization that takes into account the subjective knowledge of the agent 
is necessary. 
The rest of the paper is organized as follows. Section \ref{se:orth} describes
the main assumptions of orthodox economics and why they provide an ill-founded
basis. Section \ref{se:pre}addresses the issue of predictability, and it is
argued that the idea of having predicting model in non deterministic physics,
entails a wrong understanding of the ill-posed nature of the inverse problem.
Section \ref{se:cis} provides a new theoretical framework for modeling complex
economic systems that emphasize the relevance of adopting an ``inverse
thinking'' approach in solving the inverse problem. We conclude with Discussion
in \ref{se:di}.

\section{Orthodox economics}
\label{se:orth}
The core tenet of
orthodox economics axiomatically states that agents are 1) perfectly rational 2)
maximizers of a function cost and 3) interact in an equilibrium market. This triad has
has shown itself to be fatally flawed.
Markets are
instruments of extraordinary efficiency in processing information, integrating the views of a
large number of agents regarding the prices of complex assets. In classical
economics, it is assumed that prices fully reflect all available and relevant
information, which is equally accessible for all agents. In this view, prices
adjust almost instantaneously to every new piece of information or 
perturbation driving each price to its new equilibrium state. 

There are several
drawbacks to this theory. First, markets are composed of heterogenous agents
with very different models, motivation and strategies. Second, the rationale
that markets are regulated by a sort of homeostatic mechanism able to drive
prices to their intrinsic values, has been disproved during financial crashes.
It might be emphasized that this view is also reductionist, in the sense that
information is ultimately and entirely reflected in one variable, the price of the stock.
Third, the idea of equilibrium is an intrinsic epistemic asset in conventional
economics. Thus, an economic explanation can be seen as finding the minimum set
of basic assumptions necessary for establishing the existence of equilibrium
which is unique and stable \cite{Kaldor:1972}.
Conventional economic models, in order to make their models workable i.e. get
the analytical solutions, entail unrealistic assumptions such as, the existence
of a global conservative law or perfect competition between agents, which are
utility maximizers that make an optimal use of information that is identically
available for all the agents.

\section{A new look to predictability}
\label{se:pre}
In several occasions, economic models have shown to be powerless in predicting
bubbles and crashes that were invariably followed by important disruptions in
economic activity and even social unrest. It might be remarked that modern
macroeconomic theory may not possibly predict crisis, because it is built upon
a corpus of theoretical assumptions in which such extreme events may not be
predicted at all \cite{Lucas:2009}. Yet it is worth reminding that the use of
terms like predictability or deterministic behavior in systems of the
extraordinary complexity of national economies is, at best, a formidable
exercise of optimism.
This statement is also valid for systems of very reduced dimensionality.
Since Lorenz \cite{Lorenz:1963} and Rössler \cite{Rossler:1976},
it has been known that chaotic behavior may occur in systems with as few as three
variables. Thus we cannot pretend to find predictability in systems which are
myriad of order of magnitude larger. Clearly, we might not ask for a new
science of economics with predictive powers in situations where predictability
is out of place. We can only expect from social scientists to predict financial
bubbles and market crashes, as much as it is expected from natural scientists
to predict earthquakes, tsunamis or virus mutations. 
  
\subsection{The inverse problem}
\label{se:inv}
The main point that we want to make here is that all the 
problems in the previous section, predicting financial bubbles, tsunamis
hits etc. are inverse problems. Inverse problems are ill-founded and that is
the reason why we are bad at predicting those critical events.To solve an
inverse problem is to infer the value of parameters of interest for a given
phenomenon, based on the direct measurement of observables. This form of
inference is ill-posed in the sense that solutions to the problem may not
exist, be multiple, and be instable, that is, small error in the measurements
lead to large differences in the solution \cite{Groetsch:1993}. In engineering
the inverse problem is to solve the inverse of the forward model's equations, that
is, given the equations that describe the system's configuration or system's
internal state $m(x)$, calculate the equations of the position and momentum of the system $y$.
 \begin{equation}
m(x)=y , x=m^{-1}(y)
 \end{equation}
Systems identification or inferring the model $m$ from the accumulation of
observations $(x_i,y_i)$ is also an inverse problem, in statistics system
identification is called regression problem.

There are strong limitations to this approach, not only technical issues like
the unrealistic assumption of linearity in order to use frequency domain
techniques, but at the phenomenological level.
First, the problem is ill-posed in the sense that there are infinitum continuous
time functions $f$, that perfectly match the sampling data, that is,
we have redundancy.
Second, the problem is unstable because small errors in the 
output function $y$ may be amplified, resulting in  much greater errors in the
estimation of the function $m$. 
Interestingly, increasing the sampling rate of the measured function does not
solve this situation, it may indeed worsens it\cite{Ekstrom:1973}. This
condition needs to be conveniently recognized, specially in the current state
of increasingly powerful measurement techniques. New strategies that aim to
quantify the uncertainty in model and data need to be
explored\cite{Brenner:2010}, \cite{Tarantola:2006}

\subsection{Dealing with the bias/variance dilemma in the inverse problem}
In any process of inference lies a fundamental problem, this is what Geman calls
the bias and variance dilemma\cite{Geman:1992}.
The error in approximating a function $f(x)$ that matches the observed data
$y$, has two components, the variance which related to the uncertainty in the
measurement of data, and the bias which is due to uncertainties in the model
space, $\varepsilon=b+v$, where $\varepsilon$ is the error, $b$ is the bias and
$v$ is the variance. The bias and variance dilemma states that in order to
minimize the error we can reduce the error of one of the components, bias or variance, but not both
of them. Thus, or we bet on variance by assuming that measurement uncertainties are
trivial, or we bet on bias by neglecting inaccuracies in the model, but we can not get along
with both.
A direct implication of this statement is that the idea of finding an optimal
solution for the inverse problem must be
abandoned.

The regression problem in statistics, which is a particular case of
inverse problem, will help us to elaborate this point. The regression problem
is to find an estimator $f(x)=y$ for the purpose of approximating the desire
response $y$. The regression of $Y$ on $X$ is $E[Y|X]$, that is, the mean value
of $Y$ given $X$.
Non parametric estimators can
be arbitrarily well approximated, this property is called consistency and it is
the major reason of the popularity of non parametric estimators such as neural
networks or Boltzman machines. Consistency guarantees that for a sufficiently
big training data set, non parametric estimators achieve the best possible
performance for any learning task. It is important to note that the condition
of a ``a sufficiently big training data set' entails that consistency is an
asymptotic property, which can be formally stated as follows: 
\begin{equation}
\lim_{n\to\infty} E[f(x)-E[y|x]]=0
\end{equation}
that means that non parametric estimators $f(x)$ are consistent for all
regression problems $E[y|x]$.
But there is a toll to pay here. The versatility of the estimator to optimally approximate any task
is necessarily sensitive to the characteristics of the data. For example, 
when data samples are small or have dispersed distribution, parametric
estimators may outperform non parametric ones.
Indeed, non parametric estimators are
optimal because they are consistent, but consistency is an asymptotic property.
In real problems, the training data set can not be assumed to be arbitrarily
big, therefore we have to deal with the variance.
If we acknowledge this basic fact we can see clearly that the traditional
approach in the inverse problem, consisting on finding the operator or model that
 optimally predicts the outcomes is unrealistic because it is based on an 
 asymptotic property i.e. consistency, and technically unsuitable due to 
 nonlinearities in the functions to be optimized and the high
 dimensionality in space of candidates.
 
\section{Cisbioeconomics}
\label{se:cis}

We coin the term \emph{cisecobionomics} which refers to the study in biological
basis of how economic units make decisions to adapt within a ecosystem, using an
inner or subjective perspective. Here we understand economic unit as a system
with an internal representation of itself, for example living organisms (and
not only them) are economic units, ecosystem as the network of economic units
that are modeled using an internal or first person \emph{cis} approach. The
internal perspective aims at quantifying what information the economic unit has about
its world, rather than quantify the information that external observers have of the economic unit.
It ought to be remarked that the rationale in using optimal functions such as
utility or value in either brain and economic theories, must be found not only
in mathematical tractability. This approach entails an idea of the inverse
problem that is at odds with the ill-posed nature of the problem. 
We need to introduce a priors or bias, that is, knowledge of the model
parameters such that variability can be reduced without eliminating possible
solutions.

It has been proposed the the function of all nervous systems can be viewed as
``decision-making'' to promote future biological fitness \cite{Fiorillo:2008}.
In this respect, decision making in the brain can be seen as a process that
tries to minimize its uncertainty about the world. Thus the computational goal
of the brain is to anticipate the output of actions in order to minimize
uncertainty about its world. This view is compliant with Helmholtz
conceptualization of the brain as an inference machine that predicts
sensations. The use of Bayesian probability theory has additionally suggested a
complementary view in which the brain aims to optimize the probability
representation of what caused its sensory inputs. The Bayesian brain hypothesis
postulates the brain as an inferential machine, that makes predictions
about the world based on probabilistic models, that are updated according to the
sensorial information available at every moment. 
Friston has built a variant on
the Bayesian Brain wherein the brain optimizes a free-energy function that
tells the error between the brain internal representation and the true state of
the world that is being represented \cite{Friston:2006}], \cite{Friston:2010}.
The free energy principle integrates other global brain theories that share the view of the brain as an optimizing
machine. For example, in Hopfield's approach \cite{Hopfield:1982},
\cite{Amit:1992} neural network attractors mediate in cognitive processes like
concept formation and memory, and operate according to the optimization of an
overall energy function. In \cite{Malsburg:2010} the function of the brain is to optimize the
mismatch between sensory input and the predicted inputs of the model.
The free energy principle aims to unify theories like neural Darwinism, infomax
principle or Bayesian brain, which share a common assumption i.e. the brain always optimizes one
quantity, called value, expectation or free-energy, depending on whether the
approach relies upon economic, Bayesian or thermodynamic theory respectively.

Thus, we have identified a critical common theme that pervades in brain
function modeling and in economic systems modeling: there is one
quantity called by names like value, expected reward or utility
that is being maximized, or minimized in which case the quantity is called surprise, cost or
prediction error. 
In particular, the free energy is an upper bound in surprise
\begin{equation}
\textrm{surprise} = \frac {1}{\textrm{value}}
\end{equation}
 in such a way that organisms avoid undesirable surprising
states by minimizing their free energy, in order to to keep their internal
physiological state values within regions that promote their survival.

However, the idea that the complex
machinery of the brain may be reduced to the minimization of one single
quantity seems very unlikely. This is rooted on the mistaken understanding of
homeostasis as the universal physiological principle \cite{Gomez-Ramirez:2012}.
 There are other forms to achieve dynamic stability different to homeostasis
 \cite{Winfree:2001}. Friston understands the brain as an inference machine that
 is always optimizing its free energy by avoiding surprises, that is, the brain
 is constantly minimizing surprises which may be pernicious for the survival of
 the organism. 
It must result obvious that this view is reminiscent to idea of utility
maximisation as a mechanism that drives economies to equilibrium. Among other
things, financial crashes have very acutely showed that the conception of
economic equilibrium based on the mechanical analogy of a pendulum is
untenable. The benefit attained in terms of mathematical tractability by
adopting the hypothesis that economic agents achieve equilibrium by maximizing
utility, must not make us neglect basic facts. For example, the idea that
economic agents are utility maximizers is unverifiable, it is a necessary a
priori to solve the equations for a unique and stable equilibrium, that is to
say, we need a priori knowledge or bias to deal with the inverse problem.
The bias works as a selection mechanism that reduces the set of solutions
of the inverse problem. Once we have the models that result to apply a bias,
the next step is to test how well they predict. Thus, predictions allow us to
discard forward models when their prediction do not match a given criterion, but
not to solve the inverse problem \cite{Tarantola:2004}. The idea of using
optimization as a solver of the inverse problem is untenable and was described
in detail in section \label{se:inv}.

Economic agents, that
is, human beings provided with brains, value goods and services in order to
take decisions referred to those goods and actions in an attempt to forecast a
favorable outcome. But they decide so in multiple ways according to ecological
and historical contexts. Moreover, their actions have one direction that goes
from the irreversible past to the uncertain future. With this caveat in mind,
basic assumptions in the free-energy principle for the brain like ``any
self-organizing system that is at equilibrium with its environment must
minimize its free energy \cite{Friston:2010}'' must be carefully scrutinized.
Moreover, to assume ergodicity in a non dissipative systems like the brain is
hardly justifiable \cite{Friston:2009}. The same critic and rationale necessarily applies to economic systems
modeling.

\section{Discussion}
\label{se:di}
Complexity science has had a considerable success at addressing questions with a
new synthetic vision and a conceptual toolkit that orthodox approaches miss.
However, the answer for many of those questions remain unsolved. 
We sorely need to develop a theoretical framework, based on realistic scenarios
in which the plurality of internal motivations of the economic agents
(individuals, firms, institutions), help us to establish a systemic
understanding of complex socio-economic systems. 
In this paper we defend the view that
economics may be called to act as a natural bridge able to connect social and
technological aspects. This
positioning may sound extremely risky, the recent financial meltdown and the
inability of the economic models to forecast these extreme events, has done
nothing but reinforced the old motto ``economics is the dismal science''.  

The paper sets the basis of a new theoretical foundation to
address the inverse problem, that is, deduce models of function from
the behavioral analysis of the system, with a truly subjective or Bayesian 
approach. 
Predicting financial bubbles, the eruption of a volcano, or the formation of
cognitive neural networks in the brain are inverse problems, which as we know
since Hadamard's seminal work are not well-posed problems. The paper explores
the challenges that economic modeling faces and put them in perspective with recent
advances in brain function theory like the free-energy principle.
It provides a new perspective in tackling the inverse problem, adopting
a truly Bayesian internalistic view, that does not rely on searching an
unique solution to the inverse problem through maximization of functions like
utility. 
Here we adopt an ``inverse thinking'', which mainly relies on the
introduction of bias or a priori knowledge that constrain the solutions of the
inverse problem The candidate models are then tested against the data, that is,
those that do not predict data within an established criterion are discarded or falsified in Popperian parlance.
Note that this approach is different from solving the inverse
problem by calculating the optimal function that is the best match with the
given data. The paper builds on this approach to provide new insights to
complex system modeling that spring from a truly Bayesian approach to the 
inverse problem in complex system modeling.

\bibliographystyle{plain}
\bibliography{BrainEco}

\end{document}